\documentclass[12pt]{article}
\usepackage[latin1]{inputenc}
\usepackage{amsmath}  
\usepackage{english} 

\usepackage{amssymb} 
\usepackage{verbatim}  

\usepackage{fancyheadings} 
\renewcommand{\subsectionmark}[1]{}
\def\wwwr{{\rm I\!R}} 
\def\wwwn{{\rm I\!N}} 
\def\wwwh{{\rm h^{\!\!\!\!\!-}}}

\def\wwwq{{\mathchoice {\setbox0=\hbox{$\displaystyle\rm Q$}\hbox{\raise
0.15\ht0\hbox to0pt{\kern0.4\wd0\vrule height0.8\ht0\hss}\box0}}
{\setbox0=\hbox{$\textstyle\rm Q$}\hbox{\raise
0.15\ht0\hbox to0pt{\kern0.4\wd0\vrule height0.8\ht0\hss}\box0}}
{\setbox0=\hbox{$\scriptstyle\rm Q$}\hbox{\raise
0.15\ht0\hbox to0pt{\kern0.4\wd0\vrule height0.7\ht0\hss}\box0}}
{\setbox0=\hbox{$\scriptscriptstyle\rm Q$}\hbox{\raise
0.15\ht0\hbox to0pt{\kern0.4\wd0\vrule height0.7\ht0\hss}\box0}}}}

\begin{document}
\title{To the finite information content of the physically existing reality}

\author{Wolfgang Orthuber\thanks{
Klinikum der Christian-Albrechts-Universität zu Kiel, 
Klinik für Kieferorthopädie, D-24105 Kiel, email: mail@orthuber.com ; further info: http://www.orthuber.com/ }}
 
\date{}

\maketitle 
 
\begin{abstract}
\hspace{-2em}
Every physical measuring needs a finite, different from zero measurement time and provides information in form of the choice of a measurement result from all possible measurement results. If infinitely many (different) measurement results would be possible, the choice of a measurement result could deliver an infinite quantity of information. But the results of physical measurings (of finite duration) \emph{never} deliver an infinite quantity of information, they describe past, finite reality. Therefore the set of all possible measurement results \emph{a priori} is finite. In the physical reality only a finite information quantity can be processed within a finite time interval. For mathematical models whose representation requires a processing of an infinite quantity of information, for example irrational numbers, no (exact) equivalent exists in the physical reality. So mathematical calculations, which have an equivalent in physical reality, can include only rational (finitely many elementary) combinations of rational numbers. Conclusions arise from this for the foundations of mathematical physics.
\end{abstract}
 
\newpage

\setcounter{tocdepth}{4}
\tableofcontents
\newpage

\section{Introduction}
In the 19th century it was usual to assume a continuous behavior of physical nature and to use for it's description continuous functions with continuous sets as domain of definition and as range of values. These sets a priori contain infinitely many elements. Also the axioms of set theory permit the a priori existence of infinite sets and of choice functions on those sets. They were formulated at 1900 and led to several paradoxes (antinomies) from the beginning, which led to a discussion on the foundations of mathematics \cite{he} \cite{hi} \cite{fr} \cite{we} \cite{we1} \cite{we2} which also deals with the concept of existence (see below).
There were suggestions for different attempts to moderate the difficulties \cite{br} \cite{tr1}. But with that always was connected a limitation of mathematical liberty, so that the majority of mathematicians keeps on axioms which demand the a priori existence of infinite sets. This surely also because of the noteworthy successes of analytical approaches in the description of natural processes. So it's explainable that in mathematical physics the analytical working with infinite continuous number sets became a not scrutinized self-evident fact (exceptions cf. \cite{kh}), despite of the mentioned open discussion on the foundations, despite of the discovery of quantization of physical measurement results in the beginning of the 20th century. It has been a good opportunity for drawing conclusions with regard to the \emph{foundations} of mathematical physics, but "the moment was lost" \cite[S.15]{kh}. I think, concerning the \emph{physically}\footnote
{If something exists in the \emph{physical} meaning, it's already past and thus fixated and naturally restricted. This contradicts the concept of the infinite as something growing beyond all limits.} existing reality (short: physical reality) a scientific consensus is possible. Even Hilbert comes to the following result \cite[S.165, translated from german]{hi}:
\begin{quote}
\footnotesize{Now we have established the finiteness of the reality in two directions: to the infinite small and to the infinite large.}\normalsize
\end{quote}

Here it will be shown, that the finite \emph{information quantity} of every measurement result is closely connected with quantization and even with finiteness of the set of all possible measurement results. So for continuous number sets no equivalent can exist in physical reality.

\section{Finite information from physical measurement results}

\subsection{Information from choices within sets}
Sets can be created by subdivision of a totality into several components or elements. Already during creation of a set the choice of a sequence of elements or subsets is possible. Both the choice and the order of choices contains information. Every perception or every physical measuring provides information in form of the choice of a measurement result from all possible measurement results.

\subsection{Quantum physical aspects}
The quantum mechanical discoveries at the beginning of the 20th century have shown reductions of measuring precision as a matter of principle. Location and impulse of a particle for example are never simultaneously measurable with arbitrary precision. In the end this is consequence of the effect quantization, i.e. the fact, that only effect differences are measurable, which are multiples of the half effect quantum $\wwwh/2$.

\subsubsection{Reasons for continuous approaches}
At first this quantization has been undiscovered, because such small effect differences aren't relevant in case of usual macroscopic measurings: The systems to be measured most often are in a complex way composed of many parts whose mutual interaction and whose interaction with the surroundings isn't exactly known. For that reason there are many possibilities of uncontrolled influence on the measurement result so that it's variance is so great that effect differences in the order of $\wwwh$ have no significant influence on the measurement result. Therefore in case of macroscopic considerations it's justified to assume $\wwwh$ as negligible \emph{small} and to use analytical concepts.\\

\subsubsection{Finite information from measurement results because of effect quantization}
\label{QuantFin}
In atomic and subatomic physics the quantization of the effect becomes evident \cite[S.47]{me}, it's very important also from information theoretical point of view. Every information transmission means the transfer of free energy from a transmitter to a receiver. In case of a physical measuring the receiver consists of one or several sensors of the measurement equipment.\footnote
{If the absorption of energy at an object should be measured, the measuring is done indirectly: Initially the measurement equipment sends out free energy, which after interaction with the object is received again by sensors of the measurement equipment, from what the measurement information results.}
At this the energy is transferred to rest mass (in the sensors) by photons. For transmission the more energy is necessary, the shorter the measurement time is. If the measurement time is $t_m<\infty$ then every photon at least transfers the energy $\frac{\wwwh}{t_m}$. Since the available free energy is finite, only a finite number of photons is transferable to the sensors within the measurement time. At this energy quantities, whose difference is less than $\frac{\wwwh}{t_m}$, in principle aren't distinguishable \cite[S.129]{me}, i.e. every photon has only finitely many distinguishable possibilities for influencing the measurement equipment resp. the measurement result. Due to the finite number of sensors in which photons are absorbed, only a finite number of measurement results are possible. It is well known that this restriction is a matter of principle, it's also valid in case of an ideal, maximal exact measurement. So every measurement result is a choice from an a priori finite set and so has only finite information content. Of course another statement would contradict any everyday experience: Even the complete information of all measurings known by us is finite, it corresponds to the finite information quantity which can be known by us from (finite) past.\\

For clarification the reasoning now will be specified more precisely by information theoretical argumentation. Readers who are familiar with the possible pathologies of continuous probability distributions may skip \ref{InfoUndEntr}, the beginning of \ref{FinInfFinAcc} and continue on page \pageref{MeasuringAccBasis}.

\subsection{Information and entropy}
\label{InfoUndEntr}
Every measurement is an experiment whose result is the measurement result. The entropy\footnote
{The introduced entropy concept is closely connected to the one of thermodynamics, cf. \cite{po} and especially \cite{bri}.}
of an experiment $\beta$ quantifies its uncertainty and we shall call it $H(\beta)$. If $J$ is a set of indices, ${M:=\{A_k:k\in J\}}$ the set of all results of the experiment $\beta$ and $p(A_k)$ their probabilities, the entropy $H(\beta)$ is defined by
\begin{equation}
H(\beta):=- \sum_{k\in J} p(A_k) \log_2 p(A_k)
\label {eqdefH}
\end{equation}
(cf. \cite[S.59]{ja}), in which $\log_2$ means the logarithm to the base $2$.
$H(\beta)$ is nonnegative because of $\log_2 p(A_k)\le 0$. If $H(\beta)=0$, the result of the experiment $\beta$ is known in advance. A larger or smaller value of $H(\beta)$ corresponds to a larger or smaller uncertainty of the result. Now let be $\alpha$ an experiment which precedes $\beta$. The result of $\alpha$ can limit the number of possible results of $\beta$ and so reduce its uncertainty resp. entropy  $H(\beta)$. The entropy of $\beta$ after execution of $\alpha$ is called \emph{conditional} entropy and we write $H_\alpha(\beta)$ for it. If $\beta$ is independent of $\alpha$, the realization of $\alpha$ doesn't reduce the entropy of $\beta$, i.e. $H_\alpha(\beta)=H(\beta)$. If the result of $\alpha$ completely determines the result of $\beta$, the conditional entropy $H_\alpha(\beta)$ is zero. The difference
\begin{equation}
\label{eqdefI}
I(\alpha,\beta):=H(\beta)-H_\alpha(\beta)
\end{equation}

is called the \emph{quantity of information} contained in the result of $\alpha$ about the result of $\beta$ or shorter the \emph{information contained in $\alpha$ about $\beta$} (cf. \cite[S.86]{ja}). It shows, how much the realization of $\alpha$ reduces the uncertainty of $\beta$, how much we learn from the result of $\alpha$ about the result of $\beta$.

\subsubsection{Entropy of physical experiments}
Usually the result of a physical experiment is represented by a (if necessary multidimensional) vector, whose components are real numbers. Because the real numbers form a continuous ordered set, which (equipped with a metric) is a Hausdorff space, such representation implies infinitely many different possibilities for the result of the experiment. So its in (\ref{eqdefH}) defined entropy cannot have a finite value \cite[S.92]{ja}.\\

Without restriction of the generality we clarify this by example of a physical experiment $\beta$, whose result is an one-dimensional quantity which is represented by a real number $x\ge 0$ (multiplied by a unit), for example a length. We assume that $x$ is finite, i.e. that there is a number $s$ so that $s>x\ge 0$ holds. For a given set $M\subset\wwwr$ we write $p(M)$ for the probability that the result is contained in $M$.\\

Let's now suppose a continuous probability distribution of possible results within the interval $[0,s[$. Always we can find two numbers $a,b\in [0,s[$ with $a<b$ and $1/e>p([a,b[)>0$. The interval $[a,b[\subset [0,s[$ can be so small that the probability is distributed nearly equally within it. Then we can assume that for all $n\in\wwwn \backslash\{0\}$ and $k\in \{1,...,n\}$ the probability for the intervals
$$
J_k:=\left[a+(k-1) \frac{b-a}{n},\ a+k \frac{b-a}{n}\right[
$$
is nearly equal, i.e. $p(J_k)\approx \frac{p([a,b[)}{n}$ and with ${\epsilon:=\frac{p([a,b[)}{2}}$ particularly
$$
0<\frac{\epsilon}{n}=\frac{p([a,b[)}{2n}<p(J_k)
$$
holds. The function ${f:\ ]0,\infty[\to\wwwr,\ x\to - x \log_2 x}$ is strictly increasing in $]0,1/e[$ and $p(J_k)\in ]0,1/e[$. From this with (\ref{eqdefH}) we get for the entropy $H(\beta)$ of this experiment

\begin{eqnarray}
H(\beta)
&\ge& \sum_{k=1}^n \Big(- p(J_k) \log_2 p(J_k)\Big) 
> \sum_{k=1}^n \Big(-\frac{\epsilon}{n} \log_2 \frac{\epsilon}{n}\Big)
\label{eqEntrInfty}\\
&=& - \epsilon \log_2 \frac{\epsilon}{n}
= \epsilon (\log_2 n \ - \log_2 \epsilon)\ .\notag
\end{eqnarray}

Since at this $n$ can be arbitrarily large, we can't get a finite value for the entropy $H(\beta)$ of the experiment $\beta$. Such situation always arises, if we start out of the assumption that a continuous set of numbers represents the set of possible results of an experiment (cf. also \cite[S.93]{ja}). After execution of $\beta$ a number (the measurement result) $x\in [0,s[$ has the probability $1$, all others the probability $0$, so that the conditional entropy $H_\beta(\beta)$ of the $\beta$ is $H_\beta(\beta)= p({x}) \log_2 p({x})=1 \log_2 1 = 0$. Insertion into (\ref{eqdefI}) delivers the information quantity, which we receive from the execution of the experiment $\beta$:

\begin{equation}
I(\beta,\beta)=H(\beta)-H_\beta(\beta)=H(\beta)-0=H(\beta)
\label{eqInfoInfty}
\end{equation}
With $H(\beta)$ also the information quantity $I(\beta,\beta)$, which emerges from execution of $\beta$, isn't finite. In a nutshell: The measurement result (of the experiment $\beta$) has infinite information quantity.\\

But all experience from (finite) past has shown us, that measurement results (results of experiments with finite duration) always have only finite information quantity.

\subsection{Finite Information and finite measuring accuracy}
\label{FinInfFinAcc}
Usually one says that the reason for this is \emph{finite measuring accuracy}\footnote{This can also be a matter of principle because of quantum physical reasons (indefiniteness).}:\\
Physically possible (within finite time feasible) isn't the above mentioned experiment $\beta$, whose result is a number $x\in [0,s[$. Possible is at best an experiment $\alpha$ with finite measuring accuracy $\delta>0$, whose result is an interval $[x_\alpha -\delta,x_\alpha+\delta[\subset [0,s]$, so that the probability $P([x_\alpha-\delta, x_\alpha+\delta[)$, that the result $x$ of the experiment $\beta$ is in this interval, is great. Using some simplifications then can be shown, that the result of $\alpha$ contains only finite information (cf. \cite[S.92]{ja}), i.e. that the experiment $\alpha$ is physically possible.

\subsubsection{The concept "measuring accuracy" must have a basis}
\label{MeasuringAccBasis}
The problem of this reasoning is the usage of the term \emph{result $x$ of the experiment $\beta$}. This $x\in [0,s[$ is the result of an experiment which is not physically feasible, not even in the potential sense. Terms are used, which have never an equivalent in physical reality. So the basis for the argumentation is missing.\\
This problem always occurs, if the talk is of an experimental result represented by a selection from an infinite set of possible results, for instance in form of a number from a continuum: In this case the entropy and the gain of information (\ref{eqInfoInfty}) are not finite. So the experiment isn't feasible within finite time, i.e. it isn't physically possible. Therefore the conclusion in \ref{QuantFin} can be found also purely information theoretically. At this has to be considered, that also the duration of the experiment contains information and so there are also for it only finitely many possibilities.

\section{The finiteness of the set of possible measurement results}
We summarize the above results:
\subsection{Theorem}
\label{SFiniteManyResults}
Each physical experiment is completed after finite time. There are only finitely many possibilities for duration of the experiment. Each experimental result represents the choice from an a priori only finite number of possible results.\\

From this we can easily deduce useful conclusions for physical calculations.

\subsection{Indexing experimental results}
For instance an (if necessary multi-dimensional) index over all possible experimental results is possible and the sequence of the index is freely selectable (among others due to topological criteria, due to information theoretical coding depth). The simplest possibility is an one-dimensional index. If $M$ is the set of all possible (different) results of a physical experiment (an experiment of finite duration), we can write $M$ in the form $M=\{y_1,y_2,...,y_{|M|}\}$, in which $y_k$ can be vectors which respectively represent a test result.

\subsubsection{Example of a symmetrical index}
Often it is useful to consider symmetries. One can choose the index symmetrically to a single test result or a couple of test results and represent the set $M$ of all possible experimental results $y_k$ in following form:
$$
M=\{y_{-|M|+1},y_{-|M|+3},...,y_{-2},y_0,y_2,...,y_{|M|-3},y_{|M|-1}\}
$$
if $|M|$ is odd, and
$$
M=\{y_{-|M|+1},y_{-|M|+3},...,y_{-1},y_1,...,y_{|M|-3},y_{|M|-1}\}
$$
if $|M|$ is even.

\subsection{Finiteness of realistic physical calculations}
If an estimation of possible results of a physical experiment should be given, one has to consider that the information quantity both of the initial data and of every possible experimental result is finite. So with the help of a mathematical model from the initial data a probability distribution over a finite set of possible experimental results has to be calculated. Particularly each experimental result resp. each equivalent result of a calculation contains only finite information. So there is the possibility to calculate the result exactly from the initial data using only a finite number of elementary steps. We specify this now more precisely.

\subsubsection{Definition (Elementary combination)}
All permitted combinations of rational numbers by one of the four basic arithmetical operations (i.e. addition, subtraction, multiplication, division) are called \emph{elementary combinations}.\\

So for $a,b\in \wwwq$ there are exactly the elementary combinations $a+b=b+a, a-b, b-a, ab=ba, a/b, b/a$, in the last both cases $b\not=0$ resp. $a\not=0$ is presupposed. We know that for each elementary combination within finite time an exact equivalent can exist in the physical reality (e.g. in form of a finite sequence of binary decisions).

\subsubsection{Chaining elementary combinations}
Now for $n\in\wwwn$, $a\in\wwwq\backslash\{0\}$ let's denote by $M_n(a)$ the quantity of all numbers, which can be formed from $a$ by chaining $n$ elementary combinations. $|M_n(a)|$ is finite if $n\in\wwwn$ is a predefined (finite) number. In the reverse case, if $n$ is selectable arbitrary large subsequently, there is no upper bound for $|M_n(a)|$.

The initial data of a physical experiment (of finite duration) represent (because of their finite information content) the choice from a finite number of possible initial data, likewise the end data resp. the experimental result. If the initial data are represented as numbers, which are not all equal to $0$, we can get an infinite number of possible results if we can combine them by infinitely many elementary combinations. But a priori we know that in case of a physical experiment (i.e. after predefined maximal time for the experiment) only a finite number of different possibilities of experimental results are possible and with it only a finite number of equivalent arithmetical results. So for a mathematical calculation which is conformal to physical reality there is an upper bound $\tilde n\in\wwwn$ for the count $n$ of used elementary combinations to get the result. Particularly all numbers representing it are values of rational functions of the initial data. Since the initial data are also results of experiments with finite duration, we can start out from the assumption that the numbers which represent these data are rational, if their units are chosen in simple\footnote
{This means that the definition is done without analytical models, particularly that no irrational number factors are contained in them. We otherwise have to admit numbers from a finite field extension of $\wwwq$.}
way, which we shall assume subsequently. So mathematical calculations, which have an equivalent in physical reality, can include only rational (finitely many elementary) combinations of rational numbers. At this we know because of quantum physical results, that as a rule the end data aren't determined by the initial data, i.e. they don't contain enough information for determination. So the result of the calculation will be a probability distribution of possible results. Each of them is calculated from the initial data by a finite sequence of elementary combinations. The choice of a certain sequence is done during the experiment by a finite number of decisions, so that also the probability for a certain sequence is a rational number. We summarize:

\subsubsection{Theorem (Finite number of elementary combinations)}
Let $x\in\wwwq^l$ denote the $l$-dimensional vector of the initial data of a physical experiment (with given finite duration). There are only finitely many different possibilities for the experimental result. If $y_j, j\in\{1,2,...n\}$ are the possible $m$-dimensional result vectors with the probabilities $p_j$, both $y_j$ and $p_j$ result from $x$ by a finite number of elementary combinations. Particularly they are results of rational functions of $x$.

\newpage

\end{document}